\documentclass[eqsecnum,floats,nofootinbib,amsmath,amssymb, prd]{revtex4}

\usepackage{amsfonts,amsthm,amscd, bbm}

\usepackage{graphicx}
\usepackage[colorlinks=true,
            linkcolor=black,
            urlcolor=black,
            citecolor=black]{hyperref}
\usepackage[hyperref]{xcolor}

\definecolor{darkred}{rgb}{.95,.0,.0}
\usepackage{enumerate} 
\usepackage{colordvi} 
\usepackage{units}
\usepackage{epsfig}
\usepackage{xcolor}
\usepackage[english]{babel}
\usepackage{tensor}
\usepackage{leftidx}
\usepackage[T1]{fontenc}
\usepackage{natbib}
\usepackage{mathtools}

\usepackage{tcolorbox}

\newcommand{\be}{\begin{equation}}
\newcommand{\ee}{\end{equation}}
\newcommand{\beq}{\begin{equation}}
\newcommand{\eeq}{\end{equation}}
\newcommand{\bes}{\begin{eqnarray}}
\newcommand{\ees}{\end{eqnarray}}
\newcommand{\bqa}{\begin{eqnarray}}
\newcommand{\eqa}{\end{eqnarray}}
\newcommand{\bea}{\begin{eqnarray}}
\newcommand{\eea}{\end{eqnarray}}

\newcommand{\so}{\mathfrak{so}}

\newcommand{\an}{\mathfrak{an}}

\newcommand{\cF}{{\cal F}}

\newcommand{\cT}{{\cal T}}

\newcommand{\cI}{{\cal I}}

\def\demi{{\frac{1}{2}}}

\def\Om{{{\Omega}}}

\def\mg{{\mathfrak{g}}}
\def\mh{{\mathfrak{h}}}

\usepackage{graphicx}
\usepackage{xcolor}
\definecolor{darkgreen}{rgb}{0.01, 0.75, 0.24}
\usepackage{ulem}

\def\mG{{\mathfrak{G}}}

\def\dr{{\rightarrow}}

\def\demi{{\frac{1}{2}}}

\def\nn{\nonumber}

\def\rsc{{\rhd\!\!\!<}}
\newcommand{\rd}{\mathrm{d}}

\usepackage{accents}
\usepackage{booktabs}
\usepackage{tensor}
\DeclareMathAlphabet{\bbvar}{U}{BOONDOX-ds}{m}{n}
\DeclareMathAlphabet{\bbgreek}{U}{bbold}{m}{n}

\newcommand{\eref}[1]{(\ref{#1})}

\newcommand{\di}{\mathrm{d}}
\newcommand{\ud}[3]{\tensor{#1}{^{#2}_{#3}}}
\newcommand{\du}[3]{\tensor{#1}{_{#2}^{#3}}}
\newcommand{\circtensor}[2]{\accentset{\circ}{#1}\tensor{\vphantom{#1}}{#2}}

\def\cn{{\accentset{\circ}{\nabla}}}

\begin{document}
\title{\Large Canonical transformations generated by the boundary volume:\\Unimodular and non-Abelian teleparallel gravity }
 
\author{Florian Girelli}
\affiliation{Department of Applied Mathematics, University of Waterloo, 200 University Avenue West, Waterloo, Ontario, Canada, N2L\,3G1}
\author{Abdulmajid Osumanu}
\affiliation{Department of Applied Mathematics, University of Waterloo, 200 University Avenue West, Waterloo, Ontario, Canada, N2L\,3G1}

\author{Wolfgang Wieland}
\affiliation{Austrian Academy of Sciences, Institute for Quantum Optics and Quantum Information (IQOQI), Boltzmangasse 3, 1090 Vienna, Austria}


\begin{abstract}\noindent
Recently, a new choice of variables was identified to understand how the quantum group structure appeared in three-dimensional gravity \cite{Dupuis:2020ndx}. These variables are introduced via a canonical transformation generated by a boundary term. We show that this boundary term can actually be taken to be the volume of the boundary and that the new variables can be defined in any dimension greater than three. In addition, we study the associated metric and teleparallel formalisms. The former is a variant of the Henneaux--Teitelboim model for unimodular gravity. The latter provides a non-Abelian generalization of the usual Abelian teleparallel formulation. 
\end{abstract}


\maketitle

\tableofcontents
\hypersetup{
  linkcolor=black,
  urlcolor=darkred,
  citecolor=darkred,
}
\section{Introduction}
\noindent Quantum states are  built from the representations of the symmetries, identified at the classical level. The {structure} 
 of the symmetries depends on the choice of variables used to describe the phase space of the classical theory. In the gravity case, moving {from} 
  the metric to the frame field variables (Palatini formalism) enlarges the set of gauge symmetries since internal Lorentz rotations are absent in the usual metric formalism. On the other hand, treating gravity as a constrained topological field theory \cite{Freidel:1999rr,Freidel:2015gpa} has the opposite effect: the topological theory has a larger symmetry algebra. Gravity is recovered by imposing constraints that break the gauge symmetries down to those that appear in general relativity. Using the teleparallel formulation of gravity leads to yet another different type of symmetries, representing different representations at the quantum level (possibly related by dualities). Since both the metric and teleparallel formulations can be seen as a second-order formulation of gravity, it is interesting to study the relationship between the respective symmetries depending on the choice of variables.  We refer to \cite{Freidel:2020xyx, Freidel:2020svx, Freidel:2020ayo,DePaoli:2018erh,Oliveri:2019gvm,Wieland:2020gno,Wieland:2017zkf,Wieland:2017cmf,Wieland:2018ymr} for a recent discussion about these different issues in gravity. 

\medskip 

In field theory, canonical transformations are typically generated by adding topological invariants or boundary terms to the action. Let us list some important examples in the context of gravity below (a more detailed discussion can be found in e.g.\ \cite{Freidel:2020xyx, Freidel:2020svx, Freidel:2020ayo,DePaoli:2018erh,Oliveri:2019gvm,Wieland:2020gno,Wieland:2017zkf,Wieland:2017cmf,Wieland:2018ymr} and references therein). 
\begin{itemize}
\item In any dimension, the Gibbons-Hawking-York (boundary) term generates the canonical transformation changing the polarization from the superspace momentum representation, where the wave-functionals depend on the ADM momentum $\tilde{\pi}^{ab}=\frac{1}{16\pi G}\sqrt{\det{h}}\,(K^{ab}-h^{ab}K)$, to the position representation, where the wave-functional $\Psi[h_{ab}]$ depends only on the intrinsic metric of the boundary, c.f.\ \cite{adm,kiefer,thiemann}. 
\item For gravity in four spacetime dimensions, the Holst term induces the Ashtekar--Barbero canonical transformation from the ADM-type of variables to the (complex) Ashtekar variables, which greatly simplifies the  structure of the constraints, enabling the loop quantization of gravity \cite{newvariables,Barbero1994, holst-action, thiemann}. 
\item {In three-dimensions, quantum groups play an important role in both the quantization of Chern--Simons theory \cite{Fock:1998nu} and the discrete path integral formulation \cite{TV} of gravity. Quantum groups appear as deformed gauge symmetries, where the quantum deformation is parametrized by the cosmological constant.} This is in striking contrast to the classical formulation of three-dimensional gravity in the first-order formalism, where the gauge symmetries do \textit{not} depend on the cosmological constant.  It was recently found that a canonical transformation, induced by a boundary term, could be used to have a (bulk) theory defined in terms of new variables, such that the gauge symmetries do depend \textit{now}  on the cosmological constant as well, see \cite{Dupuis:2020ndx}. This provided a direct manner to retrieve the quantum group symmetries upon quantization. 
 \end{itemize}
 
 \medskip 

It has been conjectured that quantum groups could also be relevant for quantum gravity in four dimensions \cite{Smolin:2002sz, Fairbairn:2010cp, Han:2010pz}, so one might wonder whether the canonical transformation or the boundary term that was identified in \cite{Dupuis:2020ndx} can also be  introduced in four dimensions. We show here that this is indeed the case and that there is a straightforward generalization of the boundary term in the first-order formalism. It is then natural to ask what happens to the second-order formalism and how the metric and  teleparallel formulations of gravity could be expressed in terms of the new variables. Here, we answer these questions. In the metric formalism, we have a new notion of the Christoffel symbol. The volume term that is added to the action via the cosmological constant is replaced by the divergence of a vector field, whose norm depends on the cosmological constant. In fact, we recover a variant of the Henneaux--Teitelboim model for unimodular gravity \cite{Henneaux:1989zc}.  On the other hand, in the teleparallel formalism \cite{tele}, the new variables suggest a deformed notion of torsion, which  defines a new and non-Abelian version of the standard teleparallel formulation.  

The boundary term that generates the canonical transformation that was introduced in \cite{Dupuis:2020ndx} depends on an {internal} vector $r^I$, whose norm {sets the value of} the cosmological constant. In order to not increase the number of degrees of freedom, it was assumed that the vector is an auxiliary background structure, whose field-variations vanish on the covariant phase space, i.e.\ $\delta[r^I]=0$. However, the boundary term \cite{Dupuis:2020ndx} has  a very strong resemblance with the volume term of the boundary. It seems natural, therefore, to relax the condition $\delta[r^I]=0$. In fact, we will show that if  we actually interpret the vector $r^I$ as the \textit{internal normal}  $n^I$ to the boundary, the generating functional \cite{Dupuis:2020ndx} is exactly the volume of the boundary.  In here, we show  that given this interpretation, we can allow for $\delta[n^I]\neq0$ and \textit{still} perform  the canonical transformation, \textit{without}  adding any new degrees of freedom in the bulk. 

The new formulation has some interesting relations with already existing frameworks. 
\begin{itemize}
\item The boundary term plays an important role in holographic renormalization \cite{Bianchi:2001kw, Skenderis:2002wp} so that our new variables will also be relevant to this framework. 
\item With the change of variables, the volume term that scales with the cosmological constant drops out of the action and is replaced by the divergence  of a vector field and a constraint which sets the cosmological constant as the norm of this vector field. We will see below that the resulting action defines  a variation of the Henneaux--Teitelboim model for unimodular gravity \cite{Henneaux:1989zc}. 
\end{itemize}
\medskip 

One can often see the first-order formalism for gravity as a gauge theory, so that the canonical variables are Lie algebra-valued differential forms. This is obviously realized in three dimension  in the Chern--Simons formulation. In four dimensions, the McDowell--Mansouri formulation \cite{MacDowell:1977jt} provides a similar realization. 
Depending on the value of the cosmological constant and the signature of the metric, one usually deals with a (anti-)de Sitter or Poincar\'e (or Euclidian) gauge group, where the frame field has  values in the boost or translation sector of the group. Accordingly, the (anti-)de Sitter or Poincar\'e curvature splits  into the Lorentzian curvature sector, which depends on the cosmological constant, and the translational or boost sector, which represents torsion and is independent of $\Lambda$.  Since the (anti-)de Sitter or Poincar\'e groups can be seen as the semi-cross product between the Lorentz transformations and the boosts or translations, there is a natural action of the Lorentz group on the frame field. On the other hand, there is no action of translations back onto rotations (the commutator of a translation and a rotation has no translational part). 

The canonical transformation that we will study in here amounts to choosing a different decomposition of the (anti-)de Sitter group (the Poincar\'e case can also be treated in the same way). Indeed, if the standard formulation amounts to use the Cartan decomposition, then the canonical transformation instead picks the Iwasawa decomposition. We have then a double cross product relation and a more symmetric treatment between the connection and the frame field: the connection acts on the frame field, and the frame field acts back on the connection.   As a consequence the  (anti-)de Sitter or Poincar\'e curvature can be split differently than before. We get a generalized notion of 
 torsion, which can be viewed as a (non-Abelian) curvature in the frame field sector. This is especially relevant regarding the teleparallel formulation. 
 
We finally note that the boundary term could also be added in the usual metric formulation of General Relativity. We expect that this new boundary term will induce a canonical transformation from pure metric gravity to a  version of unimodular gravity, similar to the one introduced in this paper. Since we are primarily interested in approaches to quantum gravity based on the first-order formalism, we do not explore this thread in this paper.
 
\paragraph*{\textbf{Outline}:} The article is organized as  follows. In \hyperref[sec:matched]{section \ref{sec:matched}}, we review the gauge theory framework when dealing with a matched pair (double cross product) of Lie algebras, which will conveniently set up the conventions for the rest of the paper. In  \hyperref[sec:boundary]{section \ref{sec:boundary}}, we discuss how the canonical transformation can be implemented in any dimension by the boundary term given in terms of the volume of the boundary. In  \hyperref[sec:metric]{section \ref{sec:metric}}, we determine the  metric formulation in the new variables, while in  \hyperref[sec:tele]{section \ref{sec:tele}}, we determine the dual teleparallel formulation.

\section{Gauge theory for a matched pair of Lie groups  }\label{sec:matched}

\noindent{In the following, we develop the notion of a gauge theory for a matched pair of Lie algebras. This structure generalizes the more common semi-direct product { that} can be found in the Poincar\'e case or {in} the decomposition of the Lorentz group in terms of rotations and boosts (although they do not form a Lie algebra). Matched pairs are the relevant structure which gives rise to the quantum group structure at the quantum level. For example, in the three-dimensional gravity case. A reader who would be more geometry inclined can skip this section, which nevertheless emphasizes that what is called torsion can also be viewed as a type of curvature.  }

The algebraic background behind the matched pairs can be found in \cite{Majid:1996kd}. We call a matched pair $\mG=\mg\bowtie\mh$ of Lie algebras the pair of Lie algebras $\mg$ and $\mh$, with respective Lie brackets $[\cdot,\cdot]_\mg$ and $[\cdot,\cdot]_\mh$, equipped with two bilinear maps 
$\lhd:\mh\times\mg\rightarrow \mh, (x,\alpha)\mapsto x\lhd \alpha$ and $\rhd: \mh\times\mg\rightarrow  \mg, (x,\alpha)\mapsto x\rhd \alpha$, such that $\mh$ is a right $\mg$-module and $\mg$ is a left $\mh$-module. More intuitively, $\mg$ acts from the right on $\mh$ via $\lhd$ and there is a back action $\rhd$ of $\mh$ on $\mg$.  To guarantee the Jacobian identity on $\mG$, the left and right actions must satisfy the following compatibility conditions: $\forall \alpha, \beta \in \mg, \,  x, y \in \mh$,
\begin{subequations} \label{comp1}
\begin{align}
x \rhd [\alpha, \beta]_\mg &= [x \rhd \alpha, \beta]_\mg + [\alpha, x \rhd \beta]_\mg + (x \lhd \alpha) \rhd \beta -(x \lhd \beta) \rhd \alpha,\\
[x, y]_\mh \lhd \alpha &= [x\lhd \alpha, y]_\mh +\, [x, y\lhd \alpha]_\mh +\, x \lhd (y \rhd \alpha)- y \lhd (x \rhd \alpha).
\end{align}
\end{subequations}
As a vector space, $\mg\bowtie\mh$ is isomorphic to  $\mg\oplus\mh$. If we introduce a matched basis, such that $(\alpha,x)\in\mg\oplus\mh\sim\mg\bowtie\mh$, we may express the Lie bracket of $\mG$ in terms of $[\cdot,\cdot]_\mg$, $[\cdot,\cdot]_\mh$, $\lhd$ and $\rhd$,
\be 
\big[(\alpha, x), (\beta, y)\big]\equiv \big([\alpha, \beta]_\mg + x \rhd \beta -y \rhd \alpha,\, [x, y]_\mh + x \lhd \beta -y \lhd \alpha\big), \quad \forall \alpha, \beta \in\mg, \quad x, y\in\mh.
\ee
 In particular, there is the mixed bracket 
\be 
\big[(0,x),(\beta,0)\big]\equiv [x, \beta] =  x \lhd \beta + x \rhd \beta \in \mg\oplus \mh\label{bowtie2}.
\ee 
Notice that the left and right modules can be written in terms of the Lie bracket alone,
\be\label{different action}
x \lhd \beta := [x,\beta]\big|_{{\mh}}, \quad  x \rhd \beta:= [x,\beta]\big|_{{\mg}},
\ee
where $[x,\beta]|_{{V}}$ is the projection of the Lie bracket $[\cdot,\cdot]$ on the vector space $V$. With this notation, the compatibility relations \eqref{comp1} become
\begin{subequations}\begin{align}
x \rhd [\alpha, \beta]_\mg &= \big[[x , \alpha]\big|_{{\mg}}, \beta\big]_\mg + \big[\alpha, [x , \beta]\big|_{{\mg}}\big]_\mg + \big[[x ,\alpha]\big|_{{\mh}} ,\beta\big]\big|_{{\mg}} -\big[[x , \beta]\big|_{{\mh}},  \alpha\big]\big|_{{\mg}}, \\
[x, y]_\mh \lhd \alpha &= \big[[x, \alpha]\big|_{{\mh}}, y\big]_\mh + \big[x, [y, \alpha]\big|_{{\mh}}\big]_\mh +\big[ x , [y , \alpha]\big|_{{\mg}}\big]\big|_{{\mh}}- \big[y , [x , \alpha]\big|_{{\mg}}\big]\big|_{{\mh}}.
\end{align}
\end{subequations}


\medskip

Consider then a gauge connection $\boldsymbol{A}$ for such a Lie algebra, i.e.\ a $\mG$-valued one-form on a manifold $\mathcal{M}$, and consider its different components with respect to 
the matched Lie algebra  $\mg \bowtie \mh$. This type of gauge theory was studied by Majid \cite{Majid:1996kd} at the discrete level of parallel propagators (holonomies). Below we will derive the infinitesimal picture. 

Let us denote by $\gamma$ and $h$ the  components\footnote{We will relate $(\gamma,h)$ to the spin connection $\omega$ and the frame field $e$.} of such a gauge connection with respect to the  Lie algebras $\mg$ and $\mh$.
The $\mg \bowtie \mh$ Lie algebra-valued gauge connection takes the form 
\be
\boldsymbol{A} = \gamma+ h \in (\mg\oplus \mh)\otimes \Lambda^1(M) .
\ee
 Due to the kick-back action between  $\mg$ and $\mh$, there is a non-trivial mixing between the components of the   $\mg\bowtie\mh$ basis, see  \eqref{different action}. Consider, for example, a $\mG$-valued $p$-form $\boldsymbol{B}=(\beta,b)$. Its exterior covariant derivative is
 \bes
 \rd_{\boldsymbol{A}}\boldsymbol{B}&=& \rd \beta + \rd b + [\gamma+ h,\beta + b ]= \rd \beta + \rd b + [\gamma,\beta ]_{\mg} +[\gamma,b ] +[h,b]_{\mh} +[h,\beta ] \nn\\
 &=& \left(\rd \beta  + [\gamma,\beta ]_{\mg} +[\gamma,b ]\big|_{{\mg}} + [h,\beta ]\big|_{{\mg}}\right) +\left( \rd b+[h,b ]_{\mh} +  [\gamma,b ]\big|_{{\mh}} +[h,\beta ]\big|_{{\mh}}\right).   
 \ees
For the individual component fields $(\beta,0)$ and $(0,b)$, we  have the covariant derivatives
\begin{align}
\boldsymbol{D}_{|{\mg}}\beta &=  \rd \beta  + [\gamma,\beta ]_{\mg} + [h,\beta ]\big|_{{\mg}}, \\
\boldsymbol{D}_{|{\mh}} b &=  \rd b+[\gamma,b ]\big|_{{\mh}}+[h,b ]_{\mh}.
\end{align}
Due to the kick-back actions between the two Lie algebras, the covariant derivatives  depend on both connection components $\gamma$ and $h$. 
Consider then the components of the field strength $\boldsymbol{F}$ in the directions of $\mg$ and $\mh$,
\bes \label{new curvature}
\boldsymbol{F}[\boldsymbol{A}]&=& \rd\boldsymbol{A} + \demi [\boldsymbol{A} \wedge\boldsymbol{A}] = \rd\gamma + \rd h+ \demi \big[(\gamma +  h) \wedge (\gamma+  h)\big] \nn\\
&=& \rd\gamma + \rd h+
\demi [ \gamma \wedge \gamma ]_{{\mg}}+
 [ \gamma \wedge h]\big|_{{{\mg}}} +
 [\gamma \wedge h]\big|_{{{\mh}}}  +
\demi [h \wedge  h]_{{{\mh}}}  \nn \\
&=& \left(\rd\gamma +\demi [ \gamma \wedge \gamma ]_{{\mg}} +  [ \gamma \wedge h]\big|_{{{\mg}}} 
\right) + 
\left(\rd h
+ [\gamma \wedge h]\big|_{{{\mh}}}  +\demi [h \wedge  h]_{{{\mh}}} \right) \equiv \cF + \cT.
\ees
The physical significance of splitting $\boldsymbol{F}$ into $\cF\in\mg\otimes\Lambda^2(\mathcal{M})$ and $\cT\in\mh\otimes\Lambda^2(\mathcal{M})$ will be clear below. We will identify, in fact, $\cF$ and $\cT$ with a deformed version of $\mh$-valued torsion and $\mg$-valued curvature, 
\begin{align}
\cF & = \rd\gamma +\demi [\gamma\wedge \gamma]_{\mg} + [\gamma\wedge h]\big|_{\mg}\equiv F + [\gamma\wedge h]\big|_{\mg},\label{Fdef1}\\
\cT & = \rd h + [\gamma\wedge h]\big|_{\mh} +\demi [h\wedge h]_{\mh}\equiv T + \demi[h\wedge h]_{\mh}.\label{Tdef1}
\end{align}
 Again, the curvature components depend on both connection types. 
 The Bianchi identity for each component is
\bes
0 = \rd_{\boldsymbol{A}} \boldsymbol{F} 
&=& \rd \cF + \rd \cT + \big[(\gamma +  h) \wedge (\cF + \cT)\big]\nn\\
&=& \left(\rd \cF+ [ \gamma \wedge \cF]_{\mg} \,  + [h \wedge \cF]\big|_{{\mg}}  +[\gamma\wedge \cT]\big|_{{\mg}}   \right)   + \left(\rd \cT+ [ \gamma \wedge \cT]_{\mh} \,  + [h \wedge \cF]\big|_{{\mh}}  +[\gamma\wedge \cT]\big|_{{\mh}}   \right) .
\ees



%

Let us  consider then the gauge transformations of such a $\mG$-valued connection. Let $\boldsymbol{\lambda}=(\phi,t)$ be a $\mG$-valued scalar, which  generates an infinitesimal transformation of the connection $\boldsymbol{A}=(\gamma,h)$, 
\bes
\delta_{\boldsymbol{\lambda}}[\boldsymbol{A}]&=& \rd \boldsymbol{\lambda} + [\boldsymbol{A}, \boldsymbol{\lambda}]=   \left(\rd \phi  + [\gamma,\phi]_{\mg} +[h,\phi ]\big|_{{\mg}} + [\gamma,t ]\big|_{{\mg}}\right) +\left( \rd t+[h,t ]_{\mh} +  [\gamma,t ]\big|_{{\mh}} +[h,\phi ]\big|_{{\mh}}\right)  .
\ees
For pure rotations ($t =0$) and pure translations ($\phi=0$), we find
\be\left.
\begin{array}{rlrl}
 \delta_\phi [\gamma] &= \rd \phi  + [\gamma,\phi ]_{\mg}  + [h,\phi ]\big|_{{\mg}} =  \boldsymbol{D}_{|\mg} \phi, \quad &  \delta_\phi [h] &=  [h,\phi ]\big|_{{\mh}},  \label{transf1}  \\
 \delta_t [\gamma] &=   [\gamma,t ]\big|_{{\mg}}, & \delta_t [h] &= \rd t+[h,t ]_{\mh} +  [\gamma,t ]\big|_{{\mh}}  =  \boldsymbol{D}_{|\mh} t.
 \end{array}\hspace{0.5em}\right\}
 \ee
What is nice about these transformations is that the variation of $\gamma$ or $h$ in each of the different components is either a commutator or a derivative. In other words, the two components of the curvature two-form behave completely analogous under gauge transformations. Since $\delta_{\boldsymbol{\lambda}}[\boldsymbol{F}] = [\boldsymbol{F},\boldsymbol{\lambda}]$, we obtain 
\bes
[\cF+ \cT, \phi+ t]&=& 
 [\cF, \phi]_\mg +  [ \cF, t]\big|_{{\mg}}  + [\cT,\phi]\big|_{\mg}  +  [\cT, t]_\mh + [\cF, t]\big|_{\mh}+  [\cT, \phi]\big|_{\mh} \,.
\ees
In other words,
\be\label{transf F}\left.
\begin{array}{rlrl}
\delta_\phi [\cF] &=  [\cF, \phi]_\mg  + [\cT,\phi]\big|_{\mg},\quad & \delta_\phi [\cT]&= [\cT, \phi]\big|_{\mh}  \\
 \delta_t[\cF]&=    [ \cF, t]\big|_{{\mg}}   ,  & \delta_t [\cT]&=    [\cT, t]_\mh + [\cF,t]\big|_{\mh} .
 \end{array}\hspace{0.5em}\right\}
\ee

\medskip
A simple, but non trivial, example of such a matched pair of Lie algebras is the  Poincar\'e Lie algebra. It is a semi-cross product  $\mG=\mg\rsc \mh$, where $\mg$ is either of the Lie algebra $\so(n)$ or $\so(n-1,1)$, depending on the signature of spacetime, and $\mh\sim\mathbb{R}^n$ is the Abelian Lie algebra of translations on $\mathbb{R}^n$. 
The respective generators are the generators of (Lorentz) rotations  $J_{MN}=-J_{NM}$ and translations $P_M$, which are labelled by spacetime indices  $M,N,\dots=1,\dots,n$.

In the Poincar\'e context, there is no kick-back action of $\mh$ on $\mg$ (recall $[J,P]\propto P$). Given \eqref{different action}, we thus have
\be\label{poinc}
(\phi,t)\in \mg\rsc \mh, \quad t \lhd \phi := [t,\phi]\big|_{{\mh}}, \quad  t \rhd \phi:= [t,\phi]\big|_{{\mg}} =0.
\ee
The translations commute, i.e.\ $[\cdot,\cdot]_{\mh}=0$, and the covariant derivative for the component fields $(\phi,0)$ and $(0,t)$  with values in $\mg$ or $\mh$ is given by
  \bes
&& \boldsymbol{D}_{|\mg} \phi = \rd \phi  + [\gamma,\phi ]_{\mg}, \quad 
 \boldsymbol{D}_{|\mh}t= \rd t+  [\gamma,t ]\big|_{{\mh}}.
 \ees
Consider then the components of the curvature two-form $\boldsymbol{F}$ in the rotational and translational directions  $\mg$ and $\mh$ of the Poincar\'e Lie algebra,
 \bes \label{new curvature}
\boldsymbol{F}[\boldsymbol{A}]
&=& \left(\rd\gamma +\demi [ \gamma \wedge \gamma ]_{{\mg}} \right) + 
\left(\rd h
+ [\gamma \wedge h]\big|_{{{\mh}}} \right) 
 = F+T.
\ees
We recognize the usual notion of curvature $F$ and torsion $T$, provided we identify $\gamma$ with the spin connection $\omega$ and $h$ with the frame field $e$.

The infinitesimal gauge transformations for pure rotations ($t=0$) and pure translations ($\phi=0$) are
\be
\begin{array}{rlrl}
 \delta_\phi [\gamma] &= \rd \phi  + [\gamma,\phi ]_{\mg}   =  \boldsymbol{D}_{|\mg} \phi  , &\quad  \delta_\phi[h] &=  [h,\phi ]\big|_{{\mh}},  \label{transf1}  \\
 \delta_t [\gamma] &=   0, &  \delta_t[h] & = \rd t+  [\gamma,t ]\big|_{{\mh}}  =  \boldsymbol{D}_{|\mh} t,
 \end{array}
 \ee
where $\delta_\phi[\gamma]$ is the $\mathfrak{so}$ gauge transformation for the spin connection $\omega=\gamma$, and  $\delta_t[h]$ is an infinitesimal translation of the frame field $h=e$. Finally, there are the Poincar\'e transformations of the curvature two-form
 \be\label{transf F}
\begin{array}{rlrl}
\delta_\phi[ F] &=  [F, \phi]_\mg      , &\qquad \delta_\phi [T]&= [T, \phi]\big|_{\mh}  \\
 \delta_t [F]&=    0  , & \qquad \delta_t [T]&=    [F, t]\big|_{\mh}.
 \end{array}
\ee
Although the translations $t\in\mh$ are Abelian, a general such translation will act non-trivially on the translational curvature (torsion). If we would choose, however, a connection $\gamma$ such that $F[\gamma]=0$,  both the torsion component $T$ and the rotational (Lorentz) component $F$ of the Poincar\'e connection would be translational invariant. This choice underlies the teleparallel equivalent of general relativity.

In the following, we will consider the more complicated case of $\mG=\so(n-1,1)\sim \so(n-1) \bowtie \an_{n-1}$, which is an example of a matched Lie algebra induced by the Iwasawa decomposition of the Lorentz Lie algebra $\so(n,1)$.

\section{Canonical transformation induced by the  boundary volume term}\label{sec:boundary}

\subsection{Warm up: three-dimensions}
\noindent 
Consider the three-dimensional Einstein\,--\,Cartan action  in first-order spin-connection variables with a cosmological constant $\Lambda$, together with the following boundary  terms, whose significance will become clear below,
\begin{align}\nonumber
S_{{\text{EC}}}[e,A|n] &= \frac{1}{16\pi G} \bigg[\int_{\mathcal{M}}{\varepsilon_{IJK}\, e^{K}\wedge\Big(R^{IJ}[A] - \frac{\Lambda}{3} e^I \wedge e^J\Big)}\\
&\hspace{7em}-2\int_{\partial\mathcal{M}}\Big(s\,\varepsilon_{IJK}e^K \wedge n^I\di_A n^J+\frac{\sqrt{|\Lambda|}}{2}\varepsilon_{IJK}n^Ke^I\wedge e^J\Big)\bigg].
\label{eq:3dct1}
\end{align}
The action in the bulk is a functional of the Lorentz curvature $\ud{R}{I}{J}[A]=\di\ud{A}{I}{J}+\ud{A}{I}{K}\wedge\ud{A}{K}{J}$ and the triad $e^I$. The boundary term consists of two parts. The first term is the usual Gibbons\,--\,Hawking\,--\,York boundary term for first-order spin connection variables, where $\di_A[\cdot]=\di[\cdot] +[A,\cdot]$ is the covariant exterior derivative. The internal vector field $n^I$ is constrained to lie orthogonal to the boundary, i.e.\ $n_I\varphi_{\partial\mathcal{M}}^\ast e^I=0$, where $\varphi_{\partial\mathcal{M}}^\ast:T^\ast\mathcal{M}\rightarrow T^\ast{\partial\mathcal{M}}$ is the pull-back. In addition, $n^I$ is normalized such that $s=\eta_{IJ}n^In^J=n_I n^I=\{\pm 1,0\}$, depending on whether the boundary is spacelike, timelike, or null.\footnote{In the Lorentzian case, our metric signature is $(-$$+$$+$$\dots)$. Internal spacetime indices $I,J,K,\dots$ are raised and lowered with the internal metric tensors $\eta^{IJ}$ and $\eta_{IJ}$. Total antismmetrization of indices $I_1,I_2,\dots$ is obtained via $\omega_{[I_1\dots I_n]}=\frac{1}{n!}\sum_{\sigma\in S_n}(-1)^{\sigma}\omega_{I_{\sigma(1)}\dots I_{\sigma(n)}}$.} Its orientation is such that $n^a:=n^I\du{e}{I}{a}$ is the outwardly oriented normal to the boundary. If the torsionless equation is satisfied, then the first term is the integral of the trace of the extrinsic curvature. The second term is proportional to the induced volume of the boundary.
The boundary conditions are
\begin{equation}
\ud{h}{I}{J}\varphi^\ast_{\partial\mathcal{M}}\delta [e^J]=0,\qquad \varphi^\ast_{\partial\mathcal{M}}\delta [n_Ie^I]=0,\label{bndrycond}
\end{equation}
where $\delta [\cdot]$ is a variation on the infinite-dimensional space of kinematical histories, and $h_{IJ}=-sn_In_J+\eta_{IJ}$ is the internal projector onto the boundary.

It is also useful to evaluate the action \eref{eq:3dct1} for a spin connection $\ud{A}{I}{J}$, which is torsionless (by going \textit{half-shell}). We obtain
\begin{equation}
\label{metric formulation}
S_{\text{EC}}[A,e|n]\Big|_{\di_Ae=0}=S_{\text{EH}}[g|n] = \frac{1}{16\pi G}\left[\int_{\mathcal{M}}d^3v\big(R[g]-2\Lambda\big)+2\int_{\partial\mathcal{M}}d^2v\big(s K-\sqrt{|\Lambda|}\big)\right],
\end{equation}
where $R[g]$ is the Ricci scalar for the metric $g_{ab}=e_{Ia}\ud{e}{I}{b}$, and $K=\nabla_an^a$ is the trace of the extrinsic curvature. In addition, $d^3v$ and $d^2v$  are the canonical volume elements on $\mathcal{M}$ and $\partial{\mathcal{M}}$.\footnote{The volume elements are  $p$-forms $d^3v=\frac{1}{6}\varepsilon_{IJK}e^I\wedge e^J\wedge e^K$ and $d^2v=\frac{1}{2}n^I\varepsilon_{IJK}e^J\wedge e^K$.} The volume term appears in the definition of the bulk plus boundary action \eref{metric formulation} to cancel  infrared divergencies  that would otherwise appear when the boundary $\partial\mathcal{M}$ is sent to infinity \cite{Bianchi:2001kw}.

Consider then a region $\Sigma\subset\partial\mathcal{M}$ within the boundary. The first variation of the action for given boundary conditions \eref{bndrycond} determines the pre-symplectic potential $\Theta_\Sigma^{\text{EC}}$ on the space of physical histories, i.e.\ the space of solutions to the field equations. A straightforward calculation gives
\begin{equation}
\Theta^{\text{EC}}_\Sigma(\delta) =\frac{1}{8\pi G}\int_\Sigma \Big(s\,\varepsilon_{IJ}\delta[e^I]\wedge K^J-\sqrt{|\Lambda|}\,\varepsilon_{IJ}e^I\wedge\delta[e^J]\Big)-\frac{1}{8\pi G}\oint_{\partial\Sigma}s\,\varepsilon_{IJ}e^I\delta[n^J],\label{symplpot1}
\end{equation}
where $K^I = \varphi^\ast_{\partial M}\di_A n^I$ is the extrinsic curvature (a one-form along the boundary), and $\varepsilon_{IJ}=n^K\varepsilon_{KIJ}$ is the internal area element at the boundary. Notice that the variation $\delta[n^I]$ of the internal normal only affects a corner term to the pre-symplectic potential \eref{symplpot1}. This is a consequence of the torsionless condition $\di_Ae^I=0$ and the boundary constraints $n_In^I=s$ and $\varphi^\ast_{\partial \mathcal{M}}(n_Ie^I)=0$, hence $n_I\delta[n^I]=0$ and $\varphi^\ast_{\partial\mathcal{M}}[\epsilon_{IJK}\delta[n^I]e^J\wedge K^K]=0$.

Equation \eref{symplpot1} is a manifestation of the well-known fact that the pull-back of the triad and the extrinsic curvature are conjugate variables. From the perspective of the Chern--Simons, and Ponzano--Regge quantization of three-dimensional gravity, a connection representation is more appropriate. Following \cite{thiemann}, we consider the canonical transformation, which is generated by the Gibbons--Hawking--York boundary term
\begin{equation}
\Theta_\Sigma(\delta) := \Theta^{\text{EC}}_\Sigma(\delta) - \frac{1}{8\pi G}\delta\bigg[\int_\Sigma s\,\varepsilon_{IJ}e^I\wedge K^J\bigg].
\end{equation}
Going back to \eref{eq:3dct1}, we now immediately have
\begin{equation}\label{shifted connection}
\Theta_\Sigma(\delta) = -\frac{1}{16\pi G}\int_\Sigma \varepsilon_{IJK}e^I\wedge\delta\Big[A^{JK}-2\sqrt{|\Lambda|}\,n^{[J}e^{K]}\Big]\equiv-\frac{1}{16\pi G}\int_\Sigma \varepsilon_{IJK}e^I\wedge\delta[\Omega^{JK}],
\end{equation}
where we introduced a new connection $\Omega^{IJ}$. Notice that the variation $\delta[\Omega^{IJ}]$ contains a variation $\delta[n^I]\neq 0$. Since, however, the boundary conditions \eref{bndrycond} must be satisfied, we obtain $\delta[n^I]\perp n^I$. In addition, the vector $n^a=n^I\du{e}{I}{a}$ lies orthogonal to the boundary, hence ${\delta[n^I]}\du{e}{I}{a}$ lies tangential to the boundary. Therefore, $\frac{1}{2}\varphi_{\partial\mathcal{M}}^\ast (\epsilon_{IJK}\delta[n^I]e^J\wedge e^K)=0$, such that
\begin{equation}
\frac{1}{8\pi G}\int_\Sigma \varepsilon_{IJK}e^I\wedge\delta[n^J e^K]=\frac{1}{8\pi G}\int_\Sigma \varepsilon_{IJK}e^I\wedge n^J\delta[e^K]=\frac{1}{16\pi G}\delta\left[\int_\Sigma \varepsilon_{IJK}e^I\wedge n^Je^K\right].
\end{equation}
To solve the equations of motion in terms of the new variables $(\ud{e}{I}{a},\ud{\Omega}{IJ}{a})$, it is then necessary to smoothly extend the internal vector $n^I$ into the bulk. We thus write
\begin{align}
\Omega^{IJ} & = A^{IJ} + e^{[I} p^{J]} \equiv A^{IJ}-\cI^{IJ}, \\
\cI^{IJ}&\equiv p^{[I} e^{J]} = \demi C^{IJ}{}_K e^K, \quad C^{IJ}{}_K= (p^I \delta^J_K - p^J \delta^I_K),
\end{align}
where $p^I$ is an internal Lorentz vector that satisfies the constraints
\begin{equation} \label{normalization p}
p_Ip^I = -4\Lambda,\quad p_I\big|_{\partial \mathcal{M}} = 2\sqrt{|\Lambda|}\,n_I. 
\end{equation}
The next step ahead is to write the action in terms of the new connection $\Omega^{IJ}$.  Consider first the curvature,
\begin{align}
R_{IJ}[A]
&= R_{IJ}[\Om] + \di_{\Omega}\cI_{IJ} +{\cI}_{IK} \wedge \ud{\cI}{K}{J}\nn\\
 &=R_{IJ}[\Om] + \di_{\Omega}{\cI}_{IJ} + \frac14\left( p \wedge (p_Ie_J-p_Je_I)-p_Kp^K\,e_I\wedge e_J \right),
\end{align}
where $p=p^Ie_I$ and $\rd_{\Omega}\ud{\cI}{I}{J}= \rd\ud{\cI}{I}{J}+\ud{\Omega}{I}{K} \wedge \ud{\cI}{K}{J}  +\ud{\cI}{I}{K}\wedge\ud{\Omega}{K}{J}=\di\ud{\cI}{I}{J} +\ud{[\Omega\wedge\cI]}{I}{J}$ and we used the fact that  $e^I\wedge e_{I}=0$. With the above expression,  the first term of the action \eqref{eq:3dct1} is  
\bes
\varepsilon_{JKI}\, e^I\wedge R^{JK}[A]
&=& \varepsilon_{JKI}\, e^I\wedge \left(    R^{JK}[\Om] + \rd_{\Omega} \cI^{JK} + \demi\, p^{[J}\,p \wedge  e^{K]} - \frac{p_Mp^M}{4} e^J\wedge e^K \right)\nn\\ 
&=&\varepsilon_{JKI}\, e^I\wedge  \left(    R[\Om]^{JK} + \rd_{\Omega} \cI^{JK}  - \frac{p_Mp^M}{12}\, e^J\wedge e^K\right).
\ees
Going from the first line to the second line, we used the following identity  
\be 
 \label{secondproof1}
\frac{n_Mn^M}{6} \varepsilon_{IJK}\,e^I\wedge e^{J}\wedge e^{K}=\frac{1}{2}n^I\varepsilon_{IJK}\,n\wedge e^J\wedge e^K,
\ee
for all $n^I:n^In_I=\in\{0,\pm1\}$. Returning  to the definition of the action \eref{eq:3dct1}, we obtain
\begin{align}
S_{\text{EC}}[e,A|n] &-\frac{1}{8\pi G}{\int_{\partial\mathcal{M}}}s\,\varepsilon_{JK}e^J\wedge K^K
=\frac{1}{16\pi G}\left[\int_{\mathcal{M}}{\varepsilon_{IJK}\, e^I\wedge \bigg(R^{JK}[\Omega]  +\rd_{\Omega}\cI^{JK}\bigg)} -\frac{1}{2}{\int_{\partial\mathcal{M}}}\varepsilon_{IJK}\, e^I\wedge e^J\,p^K\right]\nn\\
&=\frac{1}{16\pi G}\int_{\mathcal{M}}{\varepsilon_{IJK}\, e^I\wedge \bigg(R[\Omega]^{JK} \, +\rd_{\Omega}\cI^{JK}} +(\di_\Omega e^{[J})p^{K]}-\frac{1}{2}e^{[J}\wedge\di_\Omega p^{K]}\bigg)\nn\\
&= \frac{1}{16\pi G}\int_{\mathcal{M}}{\varepsilon_{IJK}\, e^I\wedge \bigg(R^{JK}[\Omega] \,+\demi\, e^{[J}\wedge \rd_{\Omega}p^{K]}  \bigg)} .
\label{a-ECP action 4}
\end{align}
Equation \eref{a-ECP action 4} suggests to introduce a new action,
\begin{equation}
S[e,\Omega,p]= \frac{1}{16\pi G}\int_{\mathcal{M}}{\varepsilon_{IJK}\, e^I\wedge \bigg(R[\Omega]^{JK} \,+\demi\, e^{[J}\wedge \rd_{\Omega}p^{K]}  \bigg)},\label{newactn}
\end{equation}
where $p_I$ satisfies the mass shell condition $p_Ip^I =-4\Lambda$. To obtain the equations of motion, we vary the action for fixed boundary conditions. Given the action \eref{newactn}, the appropriate boundary conditions are
\begin{equation}
p_I\big|_{\partial M} = 2\sqrt{|\Lambda|}n_I,\quad \ud{h}{[I}{K}\ud{h}{J]}{L}\varphi^\ast_{\partial\mathcal{M}}\delta[\Omega^{KL}]=0,
\end{equation}
where $\ud{h}{I}{J}$ is the projector $\ud{h}{I}{J}=-sn^In_J+\delta^I_J$, and $\varphi^\ast_{\partial\mathcal{M}}:T^\ast M\rightarrow T^\ast(\partial \mathcal{M})$ is the pull-back. The resulting equations of motion are
\begin{align}
\mathcal{F}^{IJ}&=R^{IJ}[\Omega] + e^{[I}\wedge\rd_{\Omega}p^{J]}=0,\label{EOM3-D1}\\
 \mathcal{T}^I&=\di_\Omega e^I + \frac{1}{4}\,C_{JK}{}^{I}e^J\wedge e^K=0. \label{EOM3-D2}
\end{align}
The variation with respect to $p_I$ is redundant. Taking into account that $\delta[p_Ip^I]=2p_I\delta[p^I]=0$, i.e.\ $p^I\perp\delta[p^I]$, we obtain that the variation of the action with respect to $p^I$ vanishes provided  
\begin{equation}
\frac{1}{2}\varepsilon_{IJK}\di_\Omega(e^J\wedge e^K)\propto p_I\,d^3v.
\end{equation}
Given \eref{EOM3-D2}, this condition is always satisfied since $\frac{1}{2}\varepsilon_{IJK}\di_\Omega(e^J\wedge e^K)=-\frac{1}{2}\varepsilon_{IJK}p\wedge e^J\wedge e^K=- p_I\,d^3v$.

The field equations (\ref{EOM3-D1}, \ref{EOM3-D2}) have a simple geometric meaning. They can be rearranged into a single flatness constraint for a $\so(1,2) \bowtie \an_{3}$ matched connection. Consider the $\so(1,2) \bowtie \an_{2}$ Lie algebra-valued one-form
\begin{equation}
\boldsymbol{A}_a=\frac{1}{2}\ud{\Omega}{IJ}{a}\otimes J_{IJ} + \ud{e}{I}{a}\otimes P_I,\label{connectn}
\end{equation}
where $J_{IJ}$ are the (Lorentz) generators of  $\so(1,2)$ and $P_I$ are the generators of $\an_{2}$. The commutation relations are
\begin{subequations}
\begin{align}
\big[J_{IJ},J_{I'J'}\big]&=4\,\delta^{[R}_{I}\delta^{S]}_{J}\eta_{SS'}\delta^{[S'}_{I'}\delta^{R']}_{J'}J_{RR'},\label{JJcomm}\\
\big[P^K,J_{IJ}\big]&=2\,\delta^K_{[I}P_{J]}^{\phantom{K}}+\ud{J}{K}{[I}p_{J]}^{\phantom{K}},\label{PJcomm}\\
\big[P_I,P_J\big]&=p_{[I}P_{J]}\label{PPcomm}.
\end{align}
\end{subequations}
Going back to \eref{bowtie2}, we identify the action of the right (left) module,
\begin{equation}
P^K \lhd J_{IJ} = 2\delta^K_{[I}P_{J]}^{\phantom{K}}\in\an_{3}, \quad  P^K \rhd J_{IJ}= \ud{J}{K}{[I}p_{J]}^{\phantom{K}}\in \mathfrak{so}(1,2).
\end{equation}
To introduce the curvature of this connection, consider first the following covariant derivative, defined by its action on the basis elements of the $\so(1,2) \bowtie \an_{2}$ algebra,
\begin{equation}
\boldsymbol{\di} J_{IJ}:=0, \quad \boldsymbol{\di} P_I := \frac{1}{2}\di p^K\otimes J_{KI}. \label{covdervtn}
\end{equation}
The definition \eref{covdervtn} extends naturally to all $\so(1,2) \bowtie \an_{2}$ Lie algebra-valued $p$-forms $\boldsymbol{\omega}\in \Omega^p(\mathcal{M}:\so(1,2) \bowtie \an_{2})$ via $\boldsymbol{\di}(\boldsymbol{\omega}_1+f\boldsymbol{\omega}_2)=\boldsymbol{\di}\boldsymbol{\omega}_1+f\boldsymbol{\di}\boldsymbol{\omega}_2+\di f\wedge\boldsymbol{\omega}_2$ for all $f:\mathcal{M}\rightarrow\mathbb{R}$ and $\boldsymbol{\omega}_1,\boldsymbol{\omega}_2\in \Omega^p(\mathcal{M}:\so(1,2) \bowtie \an_{2})$. Notice also that the derivative is flat, i.e.\ $\boldsymbol{\di}^2=0$.  Consider then a second such covariant derivative $\boldsymbol{D}$, which is  defined by the deformation $\boldsymbol{D}=\boldsymbol{\di}+[\boldsymbol{A},\cdot]$. Its curvature $\boldsymbol{D}^2=[\boldsymbol{F},\cdot]$  is  given by
\begin{equation}
\boldsymbol{F}[\boldsymbol{A}] = \boldsymbol{\di}\boldsymbol{A}+\frac{1}{2}[\boldsymbol{A}\wedge \boldsymbol{A}] = \frac{1}{2}\mathcal{F}^{IJ}\otimes J_{IJ}+\mathcal{T}^I\otimes P_I=0,
\end{equation}
where the deformed $\so(1,2)$-valued curvature $\mathcal{F}$ and $\an_{2}$-valued torsion $\cT$ are defined as (\ref{EOM3-D1}, \ref{EOM3-D2}). { Let us also stress that there always exists a gauge such that $\di p=0$, such that the notion of curvature and torsion $\cF,\, \cT$ would also coincide  with (\ref{Fdef1}, \ref{Tdef1})}. If the field equations (\ref{EOM3-D1}, \ref{EOM3-D2}) are satisfied, this derivative is flat, i.e.\ $\boldsymbol{F}[\boldsymbol{A}]=0$. 
%



The torsion two-form $T^I=\di_\Omega e^I$ is now sourced by the cosmological constant, see \eref{EOM3-D2}. On the other hand, the field equations for the $\so(1,2)$ Lorentz part of the curvature two-form admit solutions where the spin connection $\ud{\Omega}{I}{J}$ is flat and $p^I$ is constant, i.e.\ $\di_\Omega p^I=0$. Hence, there is some Lorentz gauge element $\ud{\Lambda}{I}{J}:\mathcal{M}\rightarrow SO(1,2)$ such that $\ud{\Omega}{I}{J}=\du{\Lambda}{K}{I}\di\ud{\Lambda}{K}{J}$. 

Given some internal vector $p^I$ on $\mathcal{M}$, such a gauge element $\ud{\Lambda}{I}{J}$ can always be found (unless there are topological obstructions). In other words, the homogenous curvature of the underlying spacetime metric $g_{ab}=\eta_{IJ}\ud{e}{I}{a}\ud{e}{J}{b}$ has been encoded into a \textit{flat} $\so(1,2)$ connection with non-vanishing $\an_{2}$-valued torsion $T^I=\di_\Omega e^I$. This was a key feature that simplified the theory at the discrete level \cite{Dupuis:2020ndx}, and this simplification will also be relevant for us to introduce a teleparallel equivalent of gravity in the case of homogeneously curved geometries that we will discuss below.

\medskip
{ 
Before going to the general case, let us compare what we have just done with respect to \cite{Dupuis:2020ndx}. There, the starting action is 
\bes 
{S}[e,A]&\equiv 
&
 \frac{1}{16\pi G} \bigg[\int_{\mathcal{M}}{\varepsilon_{IJK}\, e^{K}\wedge\Big(R^{IJ}[A] - \frac{\Lambda}{3} e^I \wedge e^J\Big)} -\demi \int_{\partial\mathcal{M}}\Big(\varepsilon_{IJK}r^Ke^I\wedge e^J\Big)\bigg]\nn\\ 
&=&\frac{1}{16\pi G}\int_{\mathcal{M}}{\varepsilon_{IJK}\, e^I\wedge \bigg(R^{JK}[\Omega] \,+\demi\, e^{[J}\wedge \rd_{\Omega}r^{K]}  \bigg)} ,
 \label{action0}
\ees
but the  internal Lorentz vector  $r^I$ is \textit{not} interpreted as the normal of the boundary. In order to not increase the number of degrees of freedom, it is assumed that $\delta[r^I]=0$. This ensures that the $\mathfrak{so}(1,3)$ connection $\ud{A}{I}{J}$ is shifted to $\tilde \Omega^{IJ}=A^{IJ}-\,r^{[I}e^{J]} $ just as in \eqref{shifted connection}. The vector $r^I$ is also normalized as in \eqref{normalization p} in order to cancel the volume term,
\begin{equation} \label{normalization r}
r_Ir^I = -4\Lambda.
\end{equation}
Without loss of generality, the vector $r$ is assumed to have only one non-zero component, which is  proportional to the square root of the (absolute value of) the cosmological constant. As a consequence, $\di r^I=0$ and the equations of motion then coincide with the generalized curvature and torsion being zero. 
We note that there is always a choice of gauge such that the normal $p^I$ can be the constant vector $r^I$, so that the two approaches are the same.  

}
\subsection{Beyond three dimensions}
\noindent In the  {previous} section, we considered gravity in three dimensions. Our next step is to generalize the construction to arbitrary spacetime dimensions $d\geq 3$. Given the $\mathfrak{so}(1,d-1)$ spin connection $\ud{A}{I}{J}$, its conjugate momentum is the gravitational $B$-field, which is a bivector-valued $(d-2)$-form,
\be
B_{IJ}[e]= \frac{1}{(d-2)!}\,\varepsilon_{IJK_1\cdots K_{d-2}}\,e^{K_1}\wedge \cdots\wedge e^{K_{d-2}},
\ee
where $e^I$ is the co-frame that defines the metric $g_{ab}=\eta_{IJ}\ud{e}{I}{a}\ud{e}{J}{b}$. Consider then the usual Einstein\,--\,Cartan action with a boundary term proportional to the volume of the boundary
\begin{align}\nonumber
S_{\text{EC}}[e,A,p] = \frac{1}{16\pi G} &\left[\int_{\mathcal{M}}\left(B_{IJ}[e]\wedge R^{IJ}[A]  - \frac{2\Lambda}{d!} \varepsilon_{I_1\dots I_d}e^{I_1}\wedge\dots\wedge e^{I_d}\right)\right.\\
&\hspace{8em}\left.-\frac{1}{(d-1)!}\int_{\partial\mathcal{M}}p^J\varepsilon_{JI_1\dots I_{d-1}}e^{I_1}\wedge\dots\wedge e^{I_{d-1}}\right],
\label{general}
\end{align}
where $R^{IJ}[A]$ is the $\mathfrak{so}(1,d-1)$-valued curvature two-form $\ud{R}{I}{J}[A]=\di\ud{A}{I}{J}+\ud{A}{I}{K}\wedge\ud{A}{K}{J}$, and $p_I$ is an internal vector (a vector-valued $0$-form). The critical points of the action are found by {imposing} the following boundary conditions,
\begin{equation}
\ud{h}{I}{K}\ud{h}{J}{L}\varphi^\ast_{\partial\mathcal{M}}\delta[A^{KL}]=0,\qquad \delta[p_Ip^I]=0,\qquad p_I\varphi^\ast_{\partial\mathcal{M}}e^I=0,\label{bndrycond2}
\end{equation}
where $\varphi^\ast_{\partial\mathcal{M}}:T^\ast \mathcal{M}\rightarrow T^\ast(\partial\mathcal{M})$ is the pull-back and $h_{IJ}=s n_In_J +h_{IJ}$ is the projector onto the boundary such that $n_In^I=s\in\{\pm 1\}$, and $\ud{h}{I}{J}\varphi^\ast_{\partial\mathcal{M}}e^J=\varphi^\ast_{\partial\mathcal{M}}e^I$. The resulting field equations are the $d$-dimensional Einstein equations for the metric $g_{ab}=e_{Ia}\ud{e}{I}{b}$ with a cosmological constant $\Lambda$.

Consider then a region $\Sigma$ within the boundary, i.e.\ $\Sigma\subset\partial\mathcal{M}$. The pre-symplectic potential is obtained from the first variation of the action \eref{general}. Taking into account the boundary conditions \eref{bndrycond2}, we obtain 
\be 
\Theta_\Sigma(\delta) =\frac{(-1)^{(d-2)}}{16\pi G}\int_{\Sigma}B_{IJ}\wedge\delta \Big[A^{IJ} - p^{[I}e^{J]}\Big]\equiv\frac{(-1)^{(d-2)}}{16\pi G}\int_{\Sigma}B_{IJ}\wedge\delta[\Omega^{IJ}].
\ee
Next, we need to express the field equations in terms of the new and shifted connection $\Omega$. This requires  smoothly extending the internal boundary vector $p_I$ into the bulk such that $p_I\varphi^\ast_{\partial\mathcal{M}}e^I=0$ is still satisfied. Given such an extension of $p_I$ from the boundary into the bulk, we introduce the shifted connection
\begin{align}
\Omega^{IJ} & = A^{IJ} + e^{[I} p^{J]} \equiv A^{IJ}-\cI^{IJ},\label{difftensor1} \\
\cI^{IJ}&\equiv p^{[I} e^{J]} = \demi C^{IJ}{}_K e^K, \quad C^{IJ}{}_K= (p^I \delta^J_K - p^J \delta^I_K).\label{difftensor2}
\end{align}
Just as in three-dimensions, we reformulate the action in terms of the new  connection. Consider first the curvature scalar,
\begin{align}
B_{IJ}[e]\wedge R^{IJ}[A]
&=B_{IJ}[e] \wedge  \left(R^{IJ}[\Om] + \rd_{\Omega} \cI^{IJ} + \ud{\cI}{I}{K}\wedge {\cI}^{KJ} \right)\nn\\ 
&=B_{IJ}[e] \wedge  \left(R^{IJ}[\Om] + \rd_{\Omega} \cI^{IJ}+ \frac{1}{2}p^{[I}p\wedge e^{J]}-\frac{p_Kp^K}{4}e^I\wedge e^J\right),
\end{align}
where $p$ denotes the one-form $p=p_Ie^I$. We then also have the identity
\begin{equation}
 \label{secondproof2}
p_Mp^M\frac{1}{d!}\varepsilon_{I_1\dots I_d}e^{I_1}\wedge\dots\wedge e^{I_d}=\frac{1}{(d-1)!}p^J\varepsilon_{JI_1\dots I_{d-1}}p\wedge e^{I_1}\dots e^{I_{d-1}}.
\end{equation}
We thus have
\begin{equation}
\frac{1}{2}B_{IJ}\wedge\left(p^I p\wedge e^J-\frac{p_Kp^K}{2}e^I\wedge e^J\right)=-\frac{p_Mp^M}{4}\frac{(d-1)(d-2)}{d!}\varepsilon_{I_1\dots I_d}e^{I_1}\wedge\dots\wedge e^{I_d}.
\label{d-constrain vector}
\end{equation}
If we impose the mass shell condition
\begin{equation}
p_Ip^I =-\frac{8\Lambda}{(d-1)(d-2)},\label{mass-shell}
\end{equation}
then the action \eref{general} simplifies
\begin{align}
S_{\text{EC}}[e,\Omega,p] 
&= \frac{1}{16\pi G}\left[\int_{\mathcal{M}}B_{IJ}\wedge \Big(R^{IJ}[\Omega]  +\di_{\Omega}\cI^{IJ}\Big) -\frac{(-1)^{(d-2)}}{d-1}\int_{\partial\mathcal{M}}B_{IJ} \wedge p^I e^J\right]\nn\\
&= \frac{1}{16\pi G}\int_{\mathcal{M}}\left[B_{IJ}\wedge R^{IJ}[\Omega] + B_{IJ}\wedge\left(\di_{\Omega}\cI^{IJ}-p^{[I}\di_\Omega e^{J]}-\frac{1}{d-1}(\di_\Omega p^{[I})\wedge e^{J]}\right)\right].
\label{a-ECP action 3}
\end{align}
Taking into account that $\mathcal{I}^{IJ}=p^{[I}e^{J]}$, one finally arrives at the expression 
\be 
\label{general1}
S_{\text{EC}}[e,\Omega,p] 
= \frac{1}{16\pi G}\int_{\mathcal{M}}\Big( B_{IJ}\wedge R^{IJ}[\Omega] -(d-2)(-1)^{d-2} B_I \wedge\rd_{\Omega}p^I \Big),
\ee 
where we introduced the vector-valued $(d-1)$-form,
\begin{equation}
B_I=\frac{1}{(d-1)!}\varepsilon_{IK_1\dots K_{d-1}}e^{K_1}\wedge\dots\wedge e^{K_{d-1}}.
\end{equation}

The Einstein equations are the saddle points of the action \eref{general1} in the space of all fields that satisfy the mass shell constraint \eref{mass-shell} and boundary conditions
\begin{equation}
\ud{h}{I}{K}\ud{h}{J}{L}\varphi^\ast_{\partial\mathcal{M}}\delta[\Omega^{KL}]=0,\qquad \varphi^\ast_{\partial\mathcal{M}}(p_Ie^I)=0.\label{bndrycond3}
\end{equation}
The variation of the frame field for boundary conditions \eref{bndrycond3} yields the curvature constraint
\begin{equation}
\frac{1}{(d-3)!}\varepsilon_{IJKL_1\dots L_{d-3}}e^{L_1}\wedge\dots\wedge e^{L_{d-3}}\wedge\left(\underbrace{R^{JK}[\Omega]+e^{[J}\wedge\di_\Omega p^{K]}}_{\mathcal{F}^{JK}}\right)=0.\label{curvcons}
\end{equation}
By varying the action  \eqref{general1} with  respect to $\Omega$ and taking into account the boundary conditions \eref{bndrycond3}, we find that the deformed torsion vanishes\footnote{Provided $e^{I_1}\wedge\dots\wedge e^{I_d}\neq 0$.}
\be 
\label{DeformedTorsion}
\mathcal{T}^I= \rd_{\Omega} e^I + \frac{1}{4}\, \du{C}{IJ}{K}\,e^{I}\wedge e^{J}=0,
\ee
where the structure constants $\du{C}{IJ}{K}$ are given in \eref{difftensor2}. Finally, we should also consider the variations of $p^I$. Taking into account the mass shell condition \eref{mass-shell}, we obtain
\begin{equation}
\di_\Omega B_I\propto p_I.
\end{equation}
This condition is satisfied once we solve for \eref{DeformedTorsion}. Indeed, \eref{DeformedTorsion} implies
\begin{equation}
\di_\Omega B_I=\frac{1}{2}\frac{1}{(d-2)!}\varepsilon_{IK_1\dots K_{d-1}}p\wedge e^{K_1}\wedge\dots e^{K_{d-1}}=\frac{d-1}{2}p_I\,d^dv.
\end{equation}

As in three dimensions, the new variables deform the torsion two-form $T^I=\di_\Omega e ^I$, which is now sourced by the cosmological  constant. In addition, we also see that a flat $\mathfrak{so}(1,d)$ connection $\ud{\Omega}{I}{J}$ solves the curvature equation \eref{curvcons} provided $\di_\Omega p^I=0$.  This observation will be relevant once we consider a lattice approach, where the field equations are solved by imposing that the connection is piecewise flat.

The nature of the boundary (spacelike, timelike, or null) depends via the mass shell condition \eref{mass-shell} on the sign of $\Lambda$. The following table summarizes the situation for both Euclidean and Lorentzian signature in arbitrary dimensions.	
\begin{center}
\begin{tabular}{p{6.5em} p{13.5em} p{17em} } 
\cmidrule[1pt]{1-3}
& {Euclidian} & {Lorentzian} \\\cmidrule[1pt]{1-3}
\textit{Flat:} $\Lambda=0$ & $p^I=0$ or $p^I$ is Grassmanian &  $p^I=0$ or $p^I$ is light-like\\\hline
\textit{AdS:} $\Lambda<0$ & $p^I$ is real & $p^I$ is space-like or \textit{imaginary} time-like \\\hline
\textit{dS:} $\Lambda>0$ &$p^I$ is {imaginary} & $p^I$ is time-like or \textit{imaginary} space-like\\ \cmidrule[1pt]{1-3}
\end{tabular}
\end{center}

\section{Metric and deformed teleparallel formulation in the new variables} 
 
\subsection{Recovering 
unimodular gravity}\label{sec:metric}
\paragraph*{\textbf{Generalized covariant derivative.}} The  Levi-Civita connection defines the unique torsionless covariant derivative that is metric compatible.  In the last section, we deformed the notion of torsion. Let us now introduce the corresponding covariant derivative on the tangent bundle, i.e.\ the corresponding metric formulation.

For any given and invertible $d$-bein $\ud{e}{I}{a}$ and given Lorentz vector $p^I$, the field equation 
 \be 
\label{DeformedTorsion2}
 \rd_{\Omega} e^I + \frac{1}{4}\, C_{JK}{}^{I}\,e^{J}\wedge e^{K}=T^I + \frac{1}{4}\, C_{JK}{}^{I}\,e^{J}\wedge e^{K}=0
\ee
has a unique solution $\circtensor{\Omega}{^I_{Ja}}[e,p]$ for the connection $\ud{\Omega}{I}{Ja}$ in terms of $\ud{e}{I}{a}$ and $p^I$. If, in fact, $\nabla_a$ denotes the usual Levi-Civita metric compatible and torsionless covariant derivative on the tensor bundle, then we find
\begin{equation}\label{difftens}
\circtensor{\Omega}{^I_{Ja}}[e,p] = \ud{e}{I}{b}\nabla_a\du{e}{J}{b}+\ud{C}{I}{JK}\ud{e}{K}{a}=\ud{e}{I}{b}\nabla_a\du{e}{J}{b}+\ud{C}{I}{Ja}.
\end{equation}
Equation \eref{difftens} suggests to introduce the difference tensor
\begin{equation}
C_{abc}= \ud{C}{I}{JK}e_{Ia}\ud{e}{J}{b}\ud{e}{K}{c}.
\end{equation}
The corresponding metric compatible covariant derivative is defined for any smooth vector field $V^a\in T\mathcal{M}$ via
\begin{equation}
\cn_aV^b= \nabla_a V^b+\ud{C}{b}{ac}V^c.\label{difftens2}
\end{equation}
This definition generalizes naturally to arbitrary tensor fields. For any two such vector fields, we may then define curvature and torsion of this new connection,
\begin{align}
\circtensor{R}{^c_{dab}}V^d&=\cn_a\cn_bV^c-\cn_b\cn_aV^c-\circtensor{T}{^d_{ab}}\cn_dV^c,\\
\circtensor{T}{^a_{bc}}U^bV^c&=U^b\cn_bV^a-V^b\cn_bU^a-[U,V]^a.\label{torsdef}
\end{align}
Inserting \eref{difftens2} back into \eref{torsdef}, we obtain the components of the torsion two-form,
\begin{equation}
\circtensor{T}{^a_{bc}}=-\frac{1}{2}\ud{C}{a}{[bc]},\quad\text{with}\quad C_{abc}=2p_{[a}g_{b]c},
\end{equation}
which is, of course, the same as \eref{DeformedTorsion2}, but now written in the standard tensor language. { Note that we can either interpret the Christoffel  symbol $\accentset{\circ}{\Gamma}{}^{a}{}_{bc} ={\Gamma}{}^{a}{}_{bc} +C{}^{a}{}_{bc} $ as having torsion in the usual sense, or it has vanishing generalized torsion $\ud{\cT}{I}{ab}=0$.}

\paragraph*{
\textbf{Recovering unimodular gravity.}}  
As emphasized for example {in} 
\cite{BeltranJimenez:2019tjy}, the gravitational force can be  encoded into various geometric entities, namely an affine connection $\ud{\tilde{\Gamma}}{a}{bc}$, non-metricity $\tilde{Q}_{abc}$, or torsion $\ud{\tilde{T}}{a}{bc}$. Our choice of variables provides a connection $\accentset{\circ}{\Gamma}{}^{a}{}_{bc}$ where $\accentset{\circ}{Q}{}^{a}{}_{bc}=0$ but $\accentset{\circ}{T}{}^{a}{}_{bc}\neq 0$. The corresponding \textit{second-order action} is obtained by going \textit{half-shell}, i.e.\ by reinserting the solution for the connection $\accentset{\circ}{\Omega}{}^{I}{}_{Ja}$ in terms of $p_I$ and $e_{Ia}$ back into the \textit{first-order} action \eref{a-ECP action 3}. A short calculation gives
\begin{equation}
\label{EH-1}
S_{\text{EH}}[g,p]:=S_{\text{EC}}\Big[e,\accentset{\circ}{\Omega}[e,p],p\Big] = \frac{1}{16\pi G}\int_{\mathcal{M}}d^dv_g\Big(g^{cb}\circtensor{R}{^a_{cab}}[g,p]+(d-2)\overset{\circ}{\nabla}_ap^a\Big),
\end{equation}
where $d^dv_g$ is the usual metrical volume element.\footnote{The volume element $d^dv_g$ is the $d$-form, whose components are the Levi-Civita tensor, i.e.\ $(d^dv_g)_{abc\dots}=\varepsilon_{abc\dots}$.}  This action resembles the Henneaux-Teitelboim model for unimodular gravity \cite{Henneaux:1989zc}. In fact, to obtain the Einstein equations, this action is to be extremized in the space of all fields $(g_{ab},p^c)$ that satisfy the mass shell condition \eref{mass-shell} that now simply reads
\begin{equation}
g_{ab}p^ap^b = -\frac{8\Lambda}{(d-1)(d-2)}.
\end{equation}
It seems now natural to relax this condition, namely by adding a Lagrange multiplier that will impose the mass-shell constraint at the dynamical level. At the same time, this also allows us to unfreeze $\Lambda$, obtaining a  version of unimodular gravity. Consider, in fact, a $(d-1)$-form $\tau$. Its exterior derivative defines a volume density $\tilde{N}=\di\tau$. Varying the action
\begin{equation}
S_{\text{\it unimod}}[g,p,{\tau}]=\frac{1}{16\pi G}\int_{\mathcal{M}}\left[d^dv_g\Big(g^{cb}\circtensor{R}{^a_{cab}}[g,p]+(d-2)\overset{\circ}{\nabla}_ap^a\Big)-\frac{1}{2}g_{ab}p^ap^b\,\di\tau\right]
\end{equation}
with respect to $\tau$, will tell us then that $g_{ab}p^ap^b$ is constant for \textit{some} $\Lambda$, while all other variations return the usual Einstein equations with respect to the deformed connection $\accentset{\circ}{\nabla}_a$. In this way, $\tau$ provides a notion of time, $p^a$ plays the role of a space-time momentum, and $\sqrt{|\Lambda|}$ represent the rest mass of $p^a$. 

\subsection{Deformed teleparallel gravity}\label{sec:tele}


\noindent The new variables deform curvature and torsion, see \eref{EOM3-D1} and \eref{EOM3-D2}. The gravity action can be encoded in the curvature when there is no torsion or vice-versa in the torsion if there is no curvature.  The latter viewpoint underlies the teleparallel framework. We want to see now how the new variables provide a non-Abelian version of teleparallelism, where the translational connection takes values in $\mathfrak{an}_{d-1}$, whereas the Lorentzian part of the $\mathfrak{so}(1,d-1)\bowtie\mathfrak{an}_{d-1}$ connection is flat. 

Consider a generic $\mathfrak{so}(1,d-1)\bowtie\mathfrak{an}_{d-1}$ connection $\boldsymbol{A}$. The space of such $\mathfrak{so}(1,d)\bowtie\mathfrak{an}_{d-1}$ connections is an affine space. Since it is an affine space, any such connection can be written as a sum of some arbitrary reference connection $\accentset{\bullet}{\boldsymbol{A}}$ and a difference tensor $\boldsymbol{\Delta}$,
\begin{equation}
\boldsymbol{A} = \accentset{\bullet}{\boldsymbol{A}}-\boldsymbol{\Delta}.
\end{equation}
With this decomposition, we have
\begin{equation}
\boldsymbol{F}[\boldsymbol{A}]=\accentset{\bullet}{\boldsymbol{F}}-\accentset{\bullet}{\boldsymbol{D}}\boldsymbol{\Delta}+\frac{1}{2}\left[\boldsymbol{\Delta}\wedge\boldsymbol{\Delta}\right],
\end{equation}
where $\accentset{\bullet}{\boldsymbol{F}}$ is the curvature of $\accentset{\bullet}{\boldsymbol{A}}$ and $\accentset{\bullet}{\boldsymbol{D}}(\cdot)=\boldsymbol{\di}(\cdot)+[\accentset{\bullet}{\boldsymbol{A}},\cdot]$ is the covariant derivative as in \eref{covdervtn}, \eref{connectn} above.

The deformed $\mathfrak{an}_{d-1}$ Lie-algebra valued equivalent of teleparallel gravity corresponds to the following choice of reference connection,
\begin{equation}
\accentset{\bullet}{\boldsymbol{A}} = \frac{1}{2}\accentset{\bullet}{\Omega}^{IJ}\otimes J_{IJ} + e^I\otimes P_I,
\end{equation}
where $J_{IJ}$ and $P_I$ are the generators of $\mathfrak{so}(1,d-1)\bowtie\mathfrak{an}_{d-1}$ with commutation relations (\ref{JJcomm}, \ref{PJcomm}, \ref{PPcomm}). Notice that $\boldsymbol{A}$  and $\accentset{\bullet}{\boldsymbol{A}}$ have the same projection onto $\mathfrak{an}_{d}$. Therefore, $\boldsymbol{\Delta}$ is a pure rotation,
\begin{equation}
\boldsymbol{\Delta}=\frac{1}{2}\Delta^{IJ}\otimes J_{IJ}.
\end{equation}
We are now left to specify the reference connection $\ud{\accentset{\bullet}{\Omega}}{I}{J}$, and we choose it to satisfy the following two conditions:
\begin{align}
\ud{R}{I}{J}[\accentset{\bullet}{\Omega}] &=\di\ud{\accentset{\bullet}{\Omega}}{I}{J} + \ud{\accentset{\bullet}{\Omega}}{I}{K}\wedge\ud{\accentset{\bullet}{\Omega}}{K}{J} =0,\label{cond1}\\
\di_\Omega p_I &= \di p_I -\ud{\accentset{\bullet}{\Omega}}{J}{I}p_J=0 \label{cond2}.
\end{align}
{Notice that there always exists a Lorentz transformation $\Lambda{}^I{}_J$ such that $p_I= \Lambda{}^J{}_I\, r_J$, where $r_J$ is a constant vector, i.e.\ $\di r_J=0$. We can then choose 
\be
\ud{\accentset{\bullet}{\Omega}}{I}{J}= (\Lambda^{-1} \di \Lambda){}^I{}_J. 
\ee
which will then solve both \eqref{cond1} and \eqref{cond2}.  }
Therefore, the curvature of the $\mathfrak{so}(1,d-1)\bowtie\mathfrak{an}_{d-1}$ connection has only a $\mathfrak{an}_{d-1}$ part,
\begin{equation}
\accentset{\bullet}{\boldsymbol{F}} =\frac{1}{2} \accentset{\bullet}{\mathcal{F}}^{IJ}\otimes J_{IJ}+\accentset{\bullet}{\mathcal{T}}^I\otimes P_I=\accentset{\bullet}{\mathcal{T}}^I\otimes P_I, 
\end{equation}
where we defined the $\mathfrak{an}_{d}$ Lie algebra-valued torsion two-form
\begin{equation}
\accentset{\bullet}{\mathcal{T}}^I = \di\raisebox{0.2em}{${}_{\accentset{\bullet}{\Omega}}$}e^I+\frac{1}{4}\du{C}{JK}{I}e^J\wedge e^K\equiv \accentset{\bullet}{\nabla}e^I+\frac{1}{4}\du{C}{JK}{I}e^J\wedge e^K.\label{torsiondeformed}
\end{equation}
If the field equation \eref{DeformedTorsion} is satisfied, i.e.\ by going \textit{half-shell}, we can express the components of the difference tensor $\Delta_{IJ}=\Delta_{IJK}e^K$ in terms of the components of the deformed $\mathfrak{an}_{d}$ Lie algebra-valued torsion,
\begin{equation}
\Delta_{IJK}=-\frac{1}{2}\left(\accentset{\bullet}{\mathcal{T}}_{IJK}+\accentset{\bullet}{\mathcal{T}}_{JKI}-\accentset{\bullet}{\mathcal{T}}_{KIJ}\right).
\end{equation}
Inserting this solution back into the first-order action \eref{general1}, we obtain
\begin{align}
S_{\text{EC}}\Big[e,\accentset{\bullet}{\Omega}+\Delta,p\Big]&=\frac{1}{16\pi G}\int_{\mathcal{M}}\bigg[B_{IJ}\wedge R^{IJ}[\accentset{\bullet}{\Omega}]-(-1)^{d-3}B_{IJK}\wedge (\di\raisebox{0.2em}{${}_{\accentset{\bullet}{\Omega}}$}e^K)\wedge\Delta^{IJ}-B_{IJ}\wedge\ud{\Delta}{I}{L}\wedge\Delta^{LJ}\nonumber\\
&\hspace{5em}-(d-2)(-1)^{d-2}B_I\wedge\di\raisebox{0.2em}{${}_{\accentset{\bullet}{\Omega}}$} p^I+(d-2)(-1)^{d-2}B_Ip_J\wedge \Delta^{IJ}\bigg]\nonumber\\
&\hspace{5em}-\frac{1}{16\pi G}\int_{\partial\mathcal{M}}(-1)^{d-2}B_{IJ}\wedge\Delta^{IJ}\nonumber\\
&=\frac{1}{16\pi G}\int_{\mathcal{M}}\bigg[B_{IJ}\wedge \mathcal{F}^{IJ}[\accentset{\bullet}{\Omega}]-(-1)^{d-3}B_{IJK}\wedge \accentset{\bullet}{\mathcal{T}}^K\wedge\Delta^{IJ}-B_{IJ}\wedge\ud{\Delta}{I}{L}\wedge\Delta^{LJ}\nonumber\\
&\hspace{5em}-B_{IJ}\wedge e^I\wedge\di\raisebox{0.2em}{${}_{\accentset{\bullet}{\Omega}}$} p^J +\frac{1}{2}(-1)^{d-3}B_{IJK}\wedge p\wedge e^K\wedge\Delta^{IJ}\\
&\hspace{5em}-(d-2)(-1)^{d-2}B_I\wedge\di\raisebox{0.2em}{${}_{\accentset{\bullet}{\Omega}}$} p^I+(d-2)(-1)^{d-2}B_Ip_J\wedge \Delta^{IJ}\bigg]\nonumber\\
&\hspace{5em}-\frac{1}{16\pi G}\int_{\partial\mathcal{M}}(-1)^{d-2}B_{IJ}\wedge\Delta^{IJ},\label{actnstep1}
\end{align}
where we performed a partial integration and used the definition of $\accentset{\bullet}{\boldsymbol{F}}=\frac{1}{2}\accentset{\bullet}{\mathcal{F}}^{IJ}\otimes J_{IJ}+\accentset{\bullet}{\mathcal{T}}^I\otimes P_I$ for a generic $\mathfrak{so}(1,d-1)\bowtie\mathfrak{an}_{d-1}$ connection $\accentset{\bullet}{\boldsymbol{A}}$, see also \eref{EOM3-D1} and \eref{EOM3-D2}. In addition,
\begin{equation}
B_{I_1\dots I_n}:=\frac{1}{(d-n)!}\varepsilon_{I_1\dots I_nK_1\dots K_{d-n}}e^{K_1}\wedge\dots\wedge e^{K_{d-n}}.
\end{equation}
If we then also use the algebraic identity
\begin{equation}
\frac{1}{n!}B_{I_1\dots I_n}\wedge e^{J_1}\wedge\dots e^{J_n} = (-1)^{n(d-n)}d^dv\,\delta^{[J_1}_{I_1}\dots\delta^{J_n]}_{I_n},
\end{equation}
then we can write the action \eref{actnstep1} in terms of contractions of $\ud{\accentset{\bullet}{\mathcal{T}}}{I}{KJ}$ alone. A straight-forward calculation gives
\begin{align}
S_{\text{EC}}\Big[e,\accentset{\bullet}{\Omega}+\Delta,p\Big]&=\frac{1}{16\pi G}\int_{\mathcal{M}}d^dv\,\bigg[\ud{\accentset{\bullet}{\mathcal{F}}}{IJ}{IJ}-\du{e}{I}{a}\accentset{\bullet}{\nabla}_ap^I+\frac{1}{2}\ud{\accentset{\bullet}{\mathcal{T}}}{K}{IJ}\ud{\accentset{\bullet}{\mathcal{T}}}{IJ}{K}-\frac{1}{4}\ud{\accentset{\bullet}{\mathcal{T}}}{K}{IJ}\du{\accentset{\bullet}{\mathcal{T}}}{K}{IJ}+\ud{\accentset{\bullet}{\mathcal{T}}}{J}{IJ}\du{\accentset{\bullet}{\mathcal{T}}}{K}{IK}\bigg]
\\
&-\frac{1}{16\pi G}\int_{\partial\mathcal{M}}(-1)^{d-2}B_{IJ}\wedge\Delta^{IJ},\label{actnstep2}
\end{align}
where $\ud{\accentset{\bullet}{\mathcal{F}}}{IJ}{KL}$ are the components of the curvature two-form $\accentset{\bullet}{\mathcal{F}}^{IJ}=\frac{1}{2}\ud{\accentset{\bullet}{\mathcal{F}}}{IJ}{KL}e^K\wedge e^L$.

Since we can always find a reference connection $\accentset{\bullet}{\Omega}^{IJ}$ such that $\accentset{\bullet}{\nabla}_ap^I=0$ and ${R}^{IJ}[\accentset{\bullet}{\Omega}]=0$, we can now introduce the following teleparallel action, which is quadratic in the $\mathfrak{an}_{d-1}$ Lie algebra-valued field strength $\accentset{\bullet}{\mathcal{T}}^I$,
\begin{equation}
S_{\text{\it tele}}[e,\accentset{\bullet}{\Omega},p] = \frac{1}{16\pi G}\int_{\mathcal{M}}d^dv\,\bigg[\frac{1}{2}\ud{\accentset{\bullet}{\mathcal{T}}}{K}{IJ}\ud{\accentset{\bullet}{\mathcal{T}}}{IJ}{K}-\frac{1}{4}\ud{\accentset{\bullet}{\mathcal{T}}}{K}{IJ}\du{\accentset{\bullet}{\mathcal{T}}}{K}{IJ}+\ud{\accentset{\bullet}{\mathcal{T}}}{J}{IJ}\du{\accentset{\bullet}{\mathcal{T}}}{K}{IK}\bigg].\label{teleactn}
\end{equation}

%
%

\section{Outlook}
\noindent We have shown in this paper how the canonical transformation induced by  the boundary volume term was similar to the canonical transformation considered in \cite{Dupuis:2020ndx} to recover the notion of quantum group for gravity in three-dimensions. The canonical transformation results essentially in a shift of the connection by the frame field,
\be
A^{IJ}\,\dr\, \Omega^{IJ}  = A^{IJ} + e^{[I} p^{J]}, 
\ee
where $p$ is either interpreted as the normal to the boundary or as a constant vector. In both cases, the norm of $p$ is constrained to be proportional to the cosmological constant.  
This is true in any dimension higher or equal than three. In two dimensions, the frame field does not appear in the  $B$-field, so the shift cannot be done by such boundary term. This can also be seen from the normalization condition $p^2 =-\frac{8\Lambda}{(d-1)(d-2)}$.

\begin{center}
\begin{figure}[t]
\label{Fig1}
\includegraphics[scale=0.9]{summary_v2}
\caption{Brief summary of the paper: First of all, we performed a change of variables going from the spin connection to a shifted connection $\Omega$. The new connection $\Omega$ depends on a boundary field $p$, whose norm determines the cosmological constant. Integrating out the internal Lorentz transformations, we obtain a new second-order metric formulation, which resembles unimodular gravity (the norm of $p_a$ is determined by the cosmological constant). A non-Abelian version of teleparallel gravity is found by introducing a specific flat reference connection $\protect\accentset{\bullet}{\Omega}$ with $\di_{\protect\accentset{\bullet}{\Omega}}p^I=0$.}
\end{figure}
\end{center}
We would like to emphasize  that the change of variables can also be performed when $\Lambda=0$. In this case, $p$ will be  a null vector in the Lorentzian case. In the Euclidean case, we can use a Grassmanian number $\theta$ to parametrize $p$ such that $\theta^2=0$.  

\medskip 

We identified the two associated second-order theories, namely the metric and the teleparallel formulations, see the figure above. The metric formalism associated to the new variables can be seen as a new version of the Henneaux-Teitelboim model for unimodular gravity. We note that the first-order action \eqref{general1} can also be interpreted as a first-order formulation for unimodular gravity, since it can be supplemented with the normalizing constraint of the vector $p$,
\bes 
S_{\text{EC}}[e,\Omega,p] 
&=& \frac{1}{16\pi G}\int_{\mathcal{M}}\Big(\Big(  \frac{1}{(d-2)!}\,\varepsilon_{IJK_1\cdots K_{d-2}}\,e^{K_1}\wedge \cdots\wedge e^{K_{d-2}} \Big)\wedge R^{IJ}[\Omega] \nn\\&& - \frac{(d-2)(-1)^{d-2}}{(d-1)!}\Big(\varepsilon_{IK_1\dots K_{d-1}}e^{K_1}\wedge\dots e^{K_{d-1}}\Big) \wedge\rd_{\Omega}p^I  -\frac{1}{2} g_{ab}p^ap^b\,\di\tau \Big)\,
\ees 
As in the metric formulation, variation with respect to $\tau$ leads to the norm of $p$ being constant, hence recovering the cosmological constant as a constant of integration.  
The teleparallel formulation, on the other hand, is now expressed in terms of a non-Abelian field strength \eref{torsiondeformed} that enters the action quadratically \eref{teleactn}. With respect to the usual Abelian teleparallel formulation, the Abelian Lie algebra $\mathbb{R}^{d}$ is deformed  into the non-abelian Lie algebra $\mathfrak{an}_{d-1}$. 

\medskip
\noindent The results of this paper summarized in Fig.  \ref{Fig1} opens a number of interesting new directions to explore. The boundary volume term plays an important role in the holographic renormalization context \cite{Bianchi:2001kw}. It would be interesting to see whether the formulation using the new variables can also be useful, in particular, its connection with unimodular gravity. 

\medskip

The boundary term is the key  to understand why quantum groups appear in three dimensions  \cite{Dupuis:2020ndx}. In the four-dimensional context it has been conjectured that they should also be relevant \cite{Smolin:2002sz, Fairbairn:2010cp, Han:2010pz}. It would then be important to study how the charge algebra is affected by adding this term to the theory. In particular, in the list of works \cite{Freidel:2020xyx, Freidel:2020svx, Freidel:2020ayo}, it was always assumed that  $\Lambda=0$. It would be worthwhile to study how these results are affected if $\Lambda\neq0$ and the boundary volume term is also considered.   

\medskip

The non-Abelian formulation of teleparallel gravity is likely to be the classical continuum counterpart of  the (deformed) dual BF vacuum \cite{Dittrich:2016typ}. It would be interesting to construct the discrete picture, which leads upon quantization to such a deformed dual BF vacuum, generalizing \cite{Delcamp:2018sef, Dupuis:2019unm}. This is work in progress. 

\medskip

The origin of this work came from studying three-dimensional gravity. Recently, a most general bulk action for three-dimensional gravity was introduced \cite{Geiller:2020edh}. It would be interesting to see how the change of variables could be of use there, either for discretizing or in recovering the different second order formalisms.  

\medskip

Finally, the canonical transformation is implemented for a constant homogenous curvature, parametrized by the cosmological constant. One might wonder whether we could  generalize the construction in the case of a varying curvature. We leave this intriguing question for later investigations.

\providecommand{\href}[2]{#2}\begingroup\raggedright\endgroup

\end{document}